# Quantifying Phase Noise Tolerance for Single-Carrier *M*-QAM Terahertz Wireless Communications with Advantages of Photonic Approaches

Bowen Liu and Takasumi Tanabe, *Member*, *IEEE*

***Abstract*—Terahertz wireless communications offer abundant untapped spectrum and are regarded as a promising playground for next-generation high-throughput links. Yet oscillator phase noise becomes the dominant impairment at such high frequencies, severely limiting the reliability of high-order QAM transmission. While photonic approaches, such as microcombs, are known to realize ultralow phase noise, the quantitative level of suppression required to sustain reliable high-order QAM transmission has not been clarified. Here, phase noise is reconstructed from measured spectra and embedded into a single-carrier link model to evaluate its impact. Distinct distortion mechanisms are identified, with slow common phase error and instantaneous phase jitter, where the latter remains as the residual impairment after carrier phase recovery. We further adopt the 3$\sigma$ error criterion, which maps residual distortions onto the constellation, providing a clear and practical indicator of system robustness. The results indicate that modest improvements in oscillator stability translate into significant BER gains without proportional power increase. These findings provide intuitive tolerance of phase noise in *M*-QAM systems and emphasize the importance of integrating low-noise photonic oscillators such as microcombs.**

## I. INTRODUCTION

THE rapid expansion of the digital economy has triggered an unprecedented demand for wireless capacity. Monthly mobile traffic in the Asia-Pacific region increased by 11 GB per user during 2022, with more than 3.1 billion new subscriptions added in the same year [1]. Yet the average wireless connection speed in this region was only 28.5 Mbps [1]. This mismatch has become a pressing challenge for global communication infrastructure. On the other hand, newly emerging applications such as artificial intelligence (AI) networks, data center interconnects, and immersive 8K live media require transmission speeds in the order of 100~1000 Gbps [2-4], which far exceed the peak capability of current 5G (20 Gbps) architectures. Alternatively, Terahertz (THz) wireless communications spanning carrier frequencies from 0.1 to 10 THz offer a promising path to bridge this gap [5, 6]. Particularly, the 300-GHz window can provide approximately 44 GHz of underutilized spectrum, over 110 times the 400-MHz bandwidth defined in 5G NR protocols [7-9]. Benefiting from the vast spectral resource, THz wireless links could potentially deliver fiber-like capacities while maintaining the flexibility of wireless access [10]. Progress in uni-travelling-carrier photodiodes (UTC-PDs) further strengthens this prospect, providing high output power and wide bandwidth that make long-distance THz wireless links increasingly feasible [11, 12].

Early demonstrations of THz wireless transmission adopted simple formats such as amplitude shift keying (ASK) [13] and on-off keying (OOK) [14], which were limited in spectral efficiency and noise resilience. This motivated a transition to phase-based schemes such as quadrature phase shift keying (QPSK) [15-17], and further to quadrature amplitude modulation (QAM) [10, 16-26]. In 2011, a 100 Gbps polarization-multiplexed 16-QAM link was achieved with a BER of $3.1\times10^{-4}$ over 1.2 m [18], while subsequent work extended single-carrier (SC) capacity and distance. Long-reach performance has also been reported, notably 850 m of free-space delivery at 60 Gbps 16-QAM [19]. Meanwhile, quieter THz-wave sources, realized through photomixing in UTC-PDs and driven by low-noise lasers such as stimulated Brillouin scattering (SBS) fiber lasers or microresonator frequency combs from silicon nitride ($Si_3N_4$) microring resonators (MRRs), have enabled the development of higher-order modulation formats. A recent demonstration achieved 220 Gbps 32-QAM across 214 m distance with pre-FEC BER of $3.9\times10^{-3}$ [20]. Another study reported 32-QAM wireless communications at 250 Gbps over 55 m [21]. However, the constellation becomes increasingly dense beyond 32-QAM; thus, the BER would rise sharply as the rotation error dominates. This effect imposes practical limits on SC symbol rate or transmission distance. Nagatsuma's team verified 64-QAM transmissions reaching 210 Gbps data rate with $3.5\times10^{-5}$ BER [22], and 252 Gbps with $2\times10^{-2}$ BER at 20-m distance [23]. Furthermore, a 64-QAM orthogonal frequency division multiplexing (OFDM) link was achieved through two polarization multiplexing carriers, achieving ~600 Gbps at 2.8 m [24]. Adopting a multiplexing strategy of four-channel four-carrier aggregation, a throughput of ~1 Tbps of 64-QAM and 100-m indoor channel transmission was also reported [25].

This work is supported by the Japan Science and Technology Agency (JST), CRONOS, Japan (JPMJCS24N7) (Corresponding author: Takasumi Tanabe).
The authors are now with the Department of Electronics and Electrical Engineering, Faculty of Science and Technology, Keio University, 3-14-1 Hiyoshi, Kohoku-ku, Yokohama 223-8522, Japan (e-mail: b.liu@phot.elec.keio.ac.jp; takasumi@elec.keio.ac.jp).

The performance of these demonstrations is summarized in Table I. To relieve the adverse effect of phase noise, whole comb modulation was proposed to allow 64-QAM and 256-QAM links in the 300-GHz band [26].

TABLE I
DEMONSTRATIONS OF TERAHERTZ WIRELESS LINKS

| **Format** | **Data Rate** (Gbps) | **BER** | **Distance** (m) | **Ref.** |
|---|---|---|---|---|
| ASK | 1 | $1.0\times10^{-9}$ | 35 | [13] |
| OOK | 50 | $9.5\times10^{-4}$ | 100 | [14] |
| QPSK | 20×8 | $4\times10^{-4}$ | 0.5 | [15] |
| QPSK | 80 | $1.5\times10^{-6}$ | 55 | [16] |
| QPSK | 120 | $1.9\times10^{-3}$ | 110 | [17] |
| 16-QAM | 100 | $3.8\times10^{-3}$ | 20 | [10] |
| 16-QAM | 100 | $3.1\times10^{-4}$ | 1.2 | [18] |
| 16-QAM | 160 | $4.5\times10^{-3}$ | 55 | [16] |
| 16-QAM | 132 | $1.2\times10^{-2}$ | 110 | [17] |
| 16-QAM | 60 | $2.5\times10^{-2}$ | 850 | [19] |
| 16-QAM | 200 | $4.1\times10^{-3}$ | 55 | [21] |
| 16-QAM | 160 | $1.0\times10^{-5}$ | 0.055 | [22] |
| 32-QAM | 220 | $3.9\times10^{-3}$ | 214 | [20] |
| 32-QAM | 250 | $2.6\times10^{-2}$ | 55 | [21] |
| 32-QAM | 200 | $6.5\times10^{-4}$ | 0.055 | [22] |
| 64-QAM | 210 | $3.5\times10^{-3}$ | 0.055 | [22] |
| 64-QAM | 252 | $2.0\times10^{-2}$ | 20 | [23] |
| 64-QAM | 150×4 | $1.7\times10^{-2}$ | 2.8 | [24] |
| 64-QAM | 72×16 | $< 2\times10^{-2}$ | 100 | [25] |

On the other hand, $Si_3N_4$ MRR based microcombs are characterized by remarkable phase noise suppression [27-29], while optical frequency division (OFD) [30, 31], self-injection locking (SIL) [32, 33], and quiet-point engineering [34, 35] can push the phase noise floor to new lows. The full on-chip integration further strengthens microcombs as a promising enabler for THz wireless communication. Yet, the precise level of phase noise suppression required to sustain reliable $M$-QAM THz transmission, as well as the underlying distortion dynamics, remains unexplored. This gap motivates a model-based study that connects measured phase noise to symbol-level distortion and establishes practical tolerance boundaries.

In this paper, we investigate the impact of oscillator phase noise on uncoded $M$-QAM SC-THz transmission. We reconstruct temporal phase noise from measured spectra and embed it into a simplified link model to quantify symbol distortion. We also identify two distinct error mechanisms, slow common phase error (CPE) and fast instantaneous jitter, and evaluate their residual impact after digital carrier phase recovery (CPR). Furthermore, we introduce a $3\sigma$ error criterion that maps tolerance boundaries onto constellations to clarify the phase-noise thresholds that define robust operating regions for high-order QAM transmission. Through this approach, we provide quantitative guidelines on phase noise tolerance and highlight the potential advantages of microcombs toward energy- and spectrum-efficient THz wireless systems.

## II. PHASE ERROR MECHANISM AND SYSTEM MODELING

### *A. Phase Noise Mechanism of Symbol Distortion*

Figure 1 explains the mechanism by which the phase noise of a THz oscillator distorts the QAM symbols. Figure 1(a) illustrates that oscillation undergoes random phase fluctuations in the time domain, resulting in an increasing temporal uncertainty. In the frequency domain, an ideal noiseless oscillator is featured with a single sharp tone at the carrier's oscillation frequency $f_{\mathrm{LO}}$. While phase noise spreads the carrier power into sidebands, thereby reducing the carrier-to-noise ratio (CNR) as depicted in Fig. 1(b). Phase noise is generally quantified as the single-sideband (SSB) ratio within a 1-Hz offset band at frequency $f_{\mathrm{offset}}$ between the noise power $P_{\mathrm{ssb}}$ and the carrier peak power $P_{\mathrm{c}}$, expressed in dBc/Hz. Different physical processes dominate distinct regions of the noise power spectrum, each with its own offset. As illustrated in Fig. 1(c), flicker frequency noise ($1/f^3$), white frequency noise ($1/f^2$), flicker phase noise ($1/f$), and white phase noise contribute to the distinct slopes in the SSB power spectrum.

$$\mathcal{L}(f)=10\log_{10}\{\frac{2Fk_{\mathrm{B}}T}{P_{\mathrm{s}}}[1+(\frac{f_{\mathrm{LO}}}{2Qf})^2](1+\frac{f_{1/f}}{f})\} \quad (1)$$

Leeson's model governs this noise dynamics [36, 37], where $k_{\mathrm{B}}$ is the Boltzmann constant, $T$ is absolute temperature, $F$ defines the noise factor, and $P_{\mathrm{s}}$ stands for the signal power through the oscillator. Particularly, the threshold frequency $f_{\mathrm{LO}}/2Q$ between the sloped region and the white floor strongly depends on the quality factor $Q$ of the oscillator, as well as the carrier's oscillation frequency $f_{\mathrm{LO}}$. For THz communications employing extremely high-frequency carriers, frequency multiplication significantly amplifies the effective noise during up-conversion. This motivates the use of high-repetition-rate microcombs to lower the threshold frequency, thereby confining the phase noise majority to the low-offset region, reducing close-in phase uncertainty.

When the baseband QAM symbols are up-converted into a noisy carrier, the oscillator phase noise is meanwhile imposed on the transmitted waveform and propagates through the radio frequency (RF) links. Figure 1(d) shows the resulting mechanism on the constellation diagram, where the received symbols deviate from their original coordinates and form dispersed clouds of symbol mapping. As is known, a global rotation of the constellation would happen, reflecting a CPE shared among all the QAM symbols. Furthermore, local spreading of symbol mapping in both radial and azimuthal directions is also induced, which respectively corresponds to intensity noise generated by the thermal agitation of the electrons and white phase noise from instantaneous phase jitter. In a multi-carrier system such as OFDM, the PM-FM-AM conversion further impairs orthogonality among subcarriers, introducing additional interferences and error components [38]. The error vector between the transmitted and received symbols defines the error vector magnitude (EVM), which provides a concise measure of overall distortion. Additionally, Fig. 1(e) presents the time- and frequency-domain characteristics of common and instantaneous phase uncertainties. The common component $\varphi_{\text{comm}}$ varies slowly, exhibiting strong time correlation, and is dominated by low-offset colored noise. While the instantaneous component $\varphi_{\text{inst}}$ fluctuates randomly without a clear time correlation, it is associated with the white noise at high-offset frequencies. These processes together define the dominant mechanisms of symbol distortion in *M*-QAM THz links.

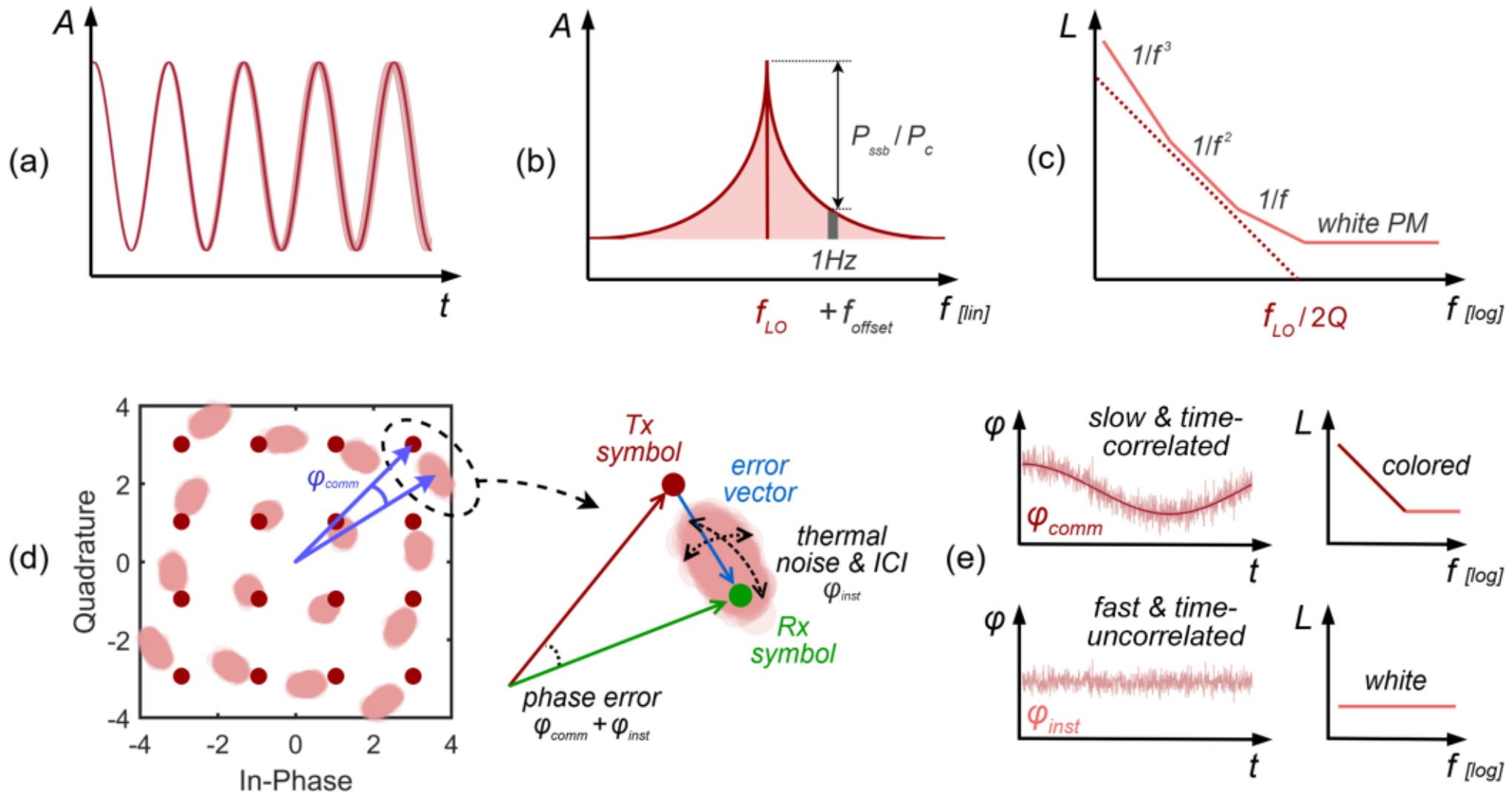

**Fig. 1.** Phase error mechanism: (a) Temporal oscillation under phase noise influence; (b) Frequency spectrum of a noisy oscillator; (c) Phase noise SSB power spectrum; (d) Phase error on the constellation diagram; (e) Low- and high-offset frequency components of the phase noise.

*B. Modeling of Terahertz Wireless Links*

The block diagram of the modeled THz wireless communication system is shown in Fig. 2(a), which consists of a digital baseband processor, an analog RF channel, and simulated THz sources. The random bit stream is modulated according to *M*-QAM mapping and then converted to an analog waveform by a digital-to-analog converter (DAC). The signal is subsequently up-converted into a 300-GHz carrier generated by the THz-wave oscillator. After translation from the baseband to the RF band, the signal waveform is shaped and transmitted by a lensed horn antenna, propagating through a standard additive white Gaussian noise (AWGN) channel. Following down-conversion and analog-to-digital conversion (ADC), digital signal processing (DSP) demaps and demodulates the received symbol sequence and estimates BER and EVM. To focus purely on the impact of phase noise in the 300-GHz band wireless transmission, the system modeling is based on the simplified situation considering a general SC link. Details not essential to phase noise, such as FEC codes and antenna patterns, are not discussed here.

The processing chain for phase noise reconstruction and symbol evaluation is explained in Fig. 2(b). Experimentally measured power spectral density (PSD) is applied as the spectra mask to initialize the phase noise characteristics of a specific oscillator in the frequency domain. Further details are then added through piecewise cubic interpolation. Its linear-scale randomization $\theta(f)$, within $[-\pi, +\pi]$, restores the stochastic nature of the noisy process [39].

$$\mathcal{O}(f)=\begin{cases}\sqrt{\dfrac{\mathcal{L}(f)}{2}}\cdot e^{j\theta(f)} & ,f>0\\ \sqrt{\dfrac{\mathcal{L}(-f)}{2}}\cdot e^{-j\theta(f)} & ,f<0\\ 1 & ,f=0\end{cases} \tag{2}$$

In this way, the time-domain realization of phase uncertainty $\varphi(n)$ is reconstructed by the inverse Fourier transform (IFFT).

$$\varphi(n)=\operatorname{Re}\{\mathcal{F}^{-1}[\mathcal{O}(f)]\} \tag{3}$$

Moreover, the communication chain is modeled in the time domain. Baseband bit stream is modulated and mapped onto an *M*-QAM constellation $u(n)$. During up-conversion, the reconstructed phase noise is imposed on the signal waveform, being transmitted through a typical AWGN channel.

$$r(n)=s(n)+n_0=u(n)\cdot e^{j\varphi(n)}+n_0 \tag{4}$$

After down-conversion and digitization, the distorted signal enters the DSP module dedicated to phase error compensation, which is implemented by a typical phase-locking loop (PLL) for CPR. The digital PLL-CPR comprises a phase-error detector, a loop filter, a direct digital synthesizer (DDS) functioning as an integration filter, and a phase shifter that applies corrections to the received symbols. The PLL can track the slow component and suppress most of the common phase jitters as,

$$y(n) = r(n) \cdot e^{j\lambda(n)} \tag{5}$$

Where $\lambda(n)$ is the output of the DDS, described by the following forward Euler integration rule with the proportional gain $g_{\mathrm{p}}$, the memorized output of the loop filter $\psi(n\text{-}1)$, and the phase error detector $e(n\text{-}1)$.

$$\lambda(n) = g_{\mathrm{p}} \cdot e(n-1) + \psi(n-1) + \lambda(n-1) \tag{6}$$

$$g_{\mathrm{p}} = \frac{4\zeta}{g_{\mathrm{e}} g_0} \cdot \frac{\kappa}{1 + 2\zeta\kappa + \kappa^2} \tag{7}$$

$$\kappa = B\tau \cdot \left( \zeta + \frac{1}{4\zeta} \right)^{-1} \tag{8}$$

For QPSK and QAM formats, the phase error detector gain $g_{\mathrm{e}}$ equals 2, and phase recovery gain $g_0$ equals sample points per symbol (here is 1). $\zeta$ is the damping factor, $B$ is the loop bandwidth, and $\tau$ is the phase-locking delay. Additionally, $\psi(n)$ and $e(n)$ are determined by the integrator gain $g_{\mathrm{i}}$, and the received symbol $r(n)$.

$$\begin{aligned} e(n) = {} & \operatorname{sgn}\left(\operatorname{Re}\{r(n)\}\right) \times \operatorname{Im}\{r(n)\} \\ & - \operatorname{sgn}\left(\operatorname{Im}\{r(n)\}\right) \times \operatorname{Re}\{r(n)\} \end{aligned} \tag{9}$$

$$\psi(n) = g_{\mathrm{i}} \cdot e(n) + \psi(n-1) \tag{10}$$

$$g_{\mathrm{i}} = \frac{4}{g_{\mathrm{e}} g_0} \cdot \frac{\kappa^2}{1 + 2\zeta\kappa + \kappa^2} \tag{11}$$

The PLL bandwidth defines the compensation range of common phase error and is set to 1% of the phase-noise sampling span (1 MHz). This parameter can be tuned according to the oscillator spectrum; for microcombs, where the white-noise corner lies at relatively low offsets, the passband can be reduced to below 10 kHz without performance degradation. After CPR, the corrected symbol sequence is sent for hard decision and to estimate root-mean-square (RMS) errors. This procedure enables a controlled and quantifiable evaluation of the phase noise impact on *M*-QAM SC-THz transmission.

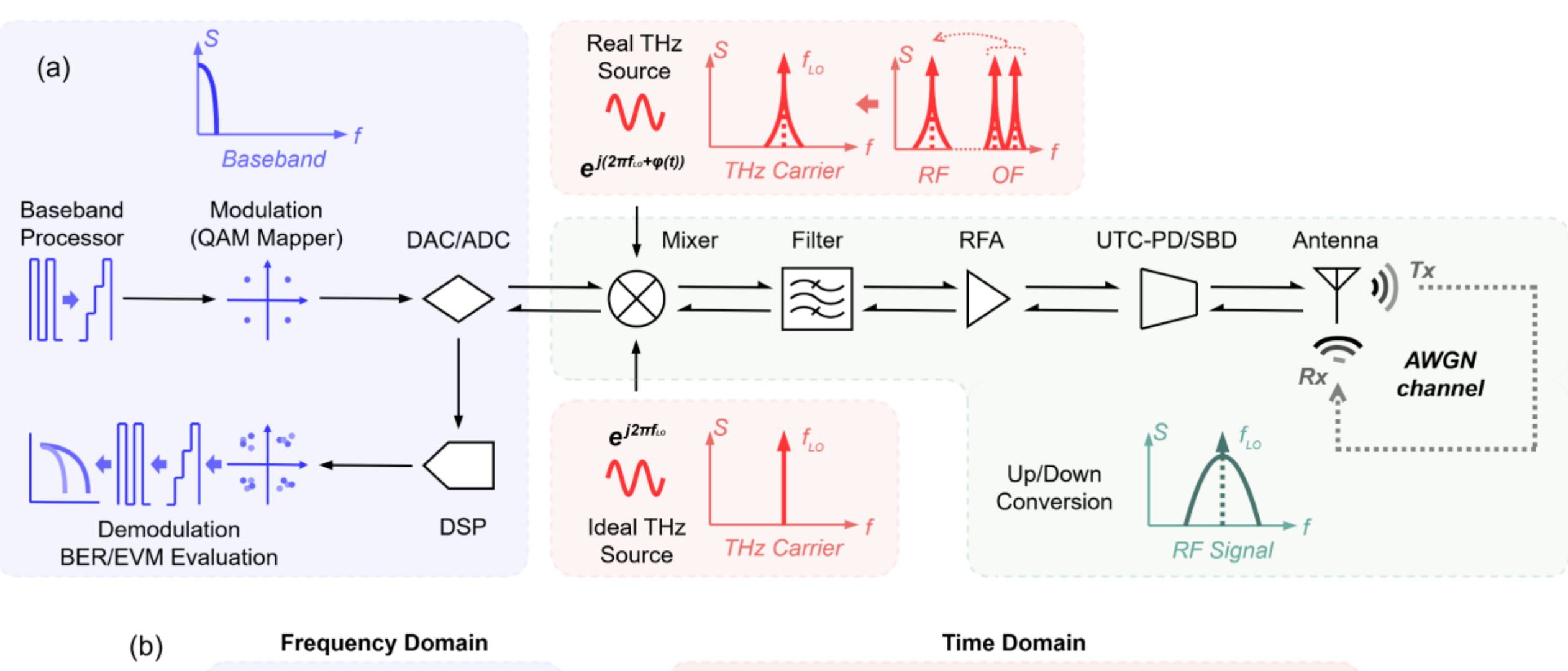

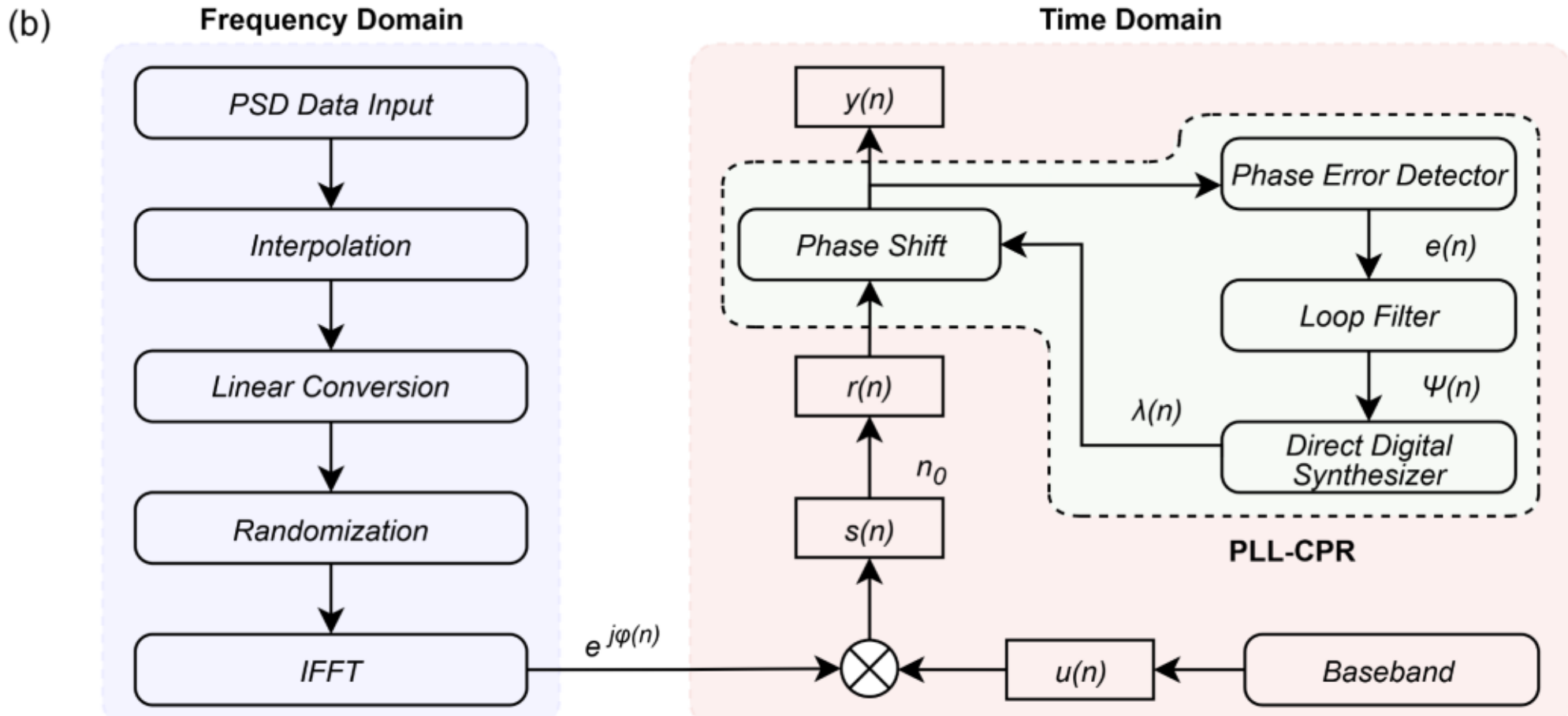

**Fig. 2.** (a) Schematic diagram of the modeled THz wireless communication system: QAM, quadrature amplitude modulation; PSK, phase shift keying; DAC, digital-to-analog converter; ADC, analog-to-digital converter; RFA, radio frequency amplifier; AWGN, additive white Gaussian noise; RF, radio frequency; OF, optical frequency; DSP, digital signal processing; BER, bit error rate; EVM, error vector magnitude. (b) Processing chain of phase noise and symbol traffic: PSD, power spectrum density; IFFT, inverse fast Fourier transform; PLL, phase-locked loop; CPR, carrier phase recovery.

## III. Results and Discussion on Phase Noise Impact

### *A. Reconstructed Phase Noise and Its Impact*

To intuitively reflect the phase noise differences, representative THz sources are analyzed, all of which are stabilized to a low noise state [8, 15, 21, 23, 31, 40-42]. Their phase noises are resampled from a bandwidth of 100 Hz to 100 MHz and then rescaled to the target carrier frequency of 300 GHz, as summarized in Table II. When the original carrier frequency is multiplied by a factor of $N$, the phase noise scales by $20{\cdot}\log_{10}N$ (in dB), which is included in the conversion. Since the variance $\sigma^2$ corresponds to the phase noise power, its standard deviation (STD) $\sigma$ provides a consistent metric for quantifying phase noise magnitude. In addition to the total STD ($\sigma_{\text{tot}}$), which includes all offset noise contributions, the instantaneous component ($\sigma_{\text{inst}}$) representing the intensity of white phase noise is extracted via a smoothing filter. This can provide a clear understanding of the residual phase noise after CPR, as well as assist in quantitatively analyzing its impact.

TABLE II
Phase Noise of Stabilized THz Oscillators Converted to 300-GHz Band

| **Oscillator** | $f_{LO}$ (GHz) | BW: $10^2 \sim 10^8$ Hz | | **Ref.** |
|---|---|---|---|---|
| | | **tot. σ** (rad) | **inst. σ** (rad) | |
| Standard VCO | 29.55 | 0.8683 | 0.2602 | [8] |
| 55-nm BiCMOS | 9.96 | 13.9318 | 0.0310 | [40] |
| 28-nm CMOS | 300 | 0.1540 | 0.1001 | [41] |
| 14-nm FinFET | 6.45 | 0.0764 | 0.0564 | [42] |
| ASG E8257D | 50 | 0.0148 | 0.0115 | [21] |
| SIL SiN Microcomb | 50 | 0.8976 | 0.0120 | [21] |
| ECL Photomixing | 350 | 0.0252 | 0.0212 | [15] |
| SBS Fiber Laser | 137.5 | 0.4295 | 0.0108 | [23] |
| 2P-OFD Microcomb | 20 | 0.0071 | 0.0058 | [31] |

Here, we adopt the five THz sources and show the reconstructed phase noise in Fig. 3. Determined by their noise floor levels, the masks in Fig. 3(a) positioned from top to bottom respectively correspond to a standard voltage-controlled oscillator (VCO) defined in 3GPP 5G NR protocols [8], a 300-GHz CMOS oscillator [41], a photonic-mixing external cavity laser (ECL) [15], a state-of-the-art arbitrary signal generator (ASG, Keysight E8257D) [21], and a microcomb stabilized with two-point optical frequency division (2P-OFD) [31]. Figure 3(b) depicts their temporal phase jitters. Distinct phase noise magnitudes and time correlations are exhibited, which are derived from the combined effect of common and instantaneous phase uncertainties ($\varphi_{\text{comm}}$ and $\varphi_{\text{inst}}$). Figure 3(c) further extracts the instantaneous phase jitters in the time domain through a smoothing filter. Microcomb, photonic-mixing ECL, and ASG schemes exhibit significantly lower noise magnitudes compared to conventional electronic oscillators. In particular, the instantaneous phase uncertainty $\varphi_{\text{inst}}$ introduced by microcombs is merely half of the ASG output and a quarter of the ECL beat signal. The corresponding probability density function (PDF) is also calculated as shown in Fig. 3(d), where normal distributions are reflected with zero expected value $\mu_{\text{inst}}$ and distinguish $\sigma_{\text{inst}}$—identifying the relative likelihood of error coverage and white phase noise magnitude.

$$f\left(\varphi \middle| \mu_{\text{inst}}, \sigma_{\text{inst}}\right) = \frac{1}{\sigma_{\text{inst}}\sqrt{2\pi}} e^{\frac{-(\varphi-\mu_{\text{inst}})^2}{2\sigma_{\text{inst}}^2}} \tag{12}$$

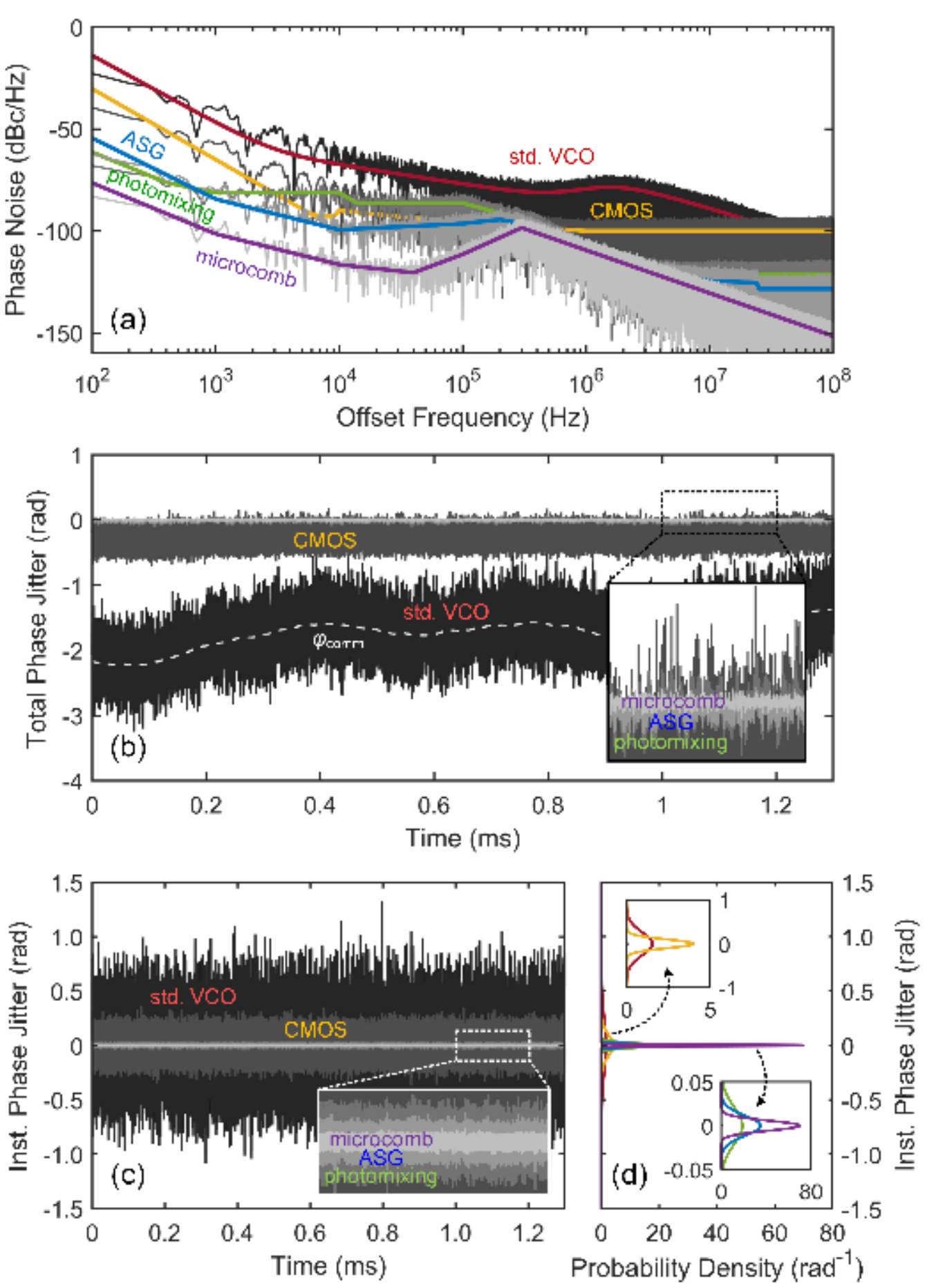

**Fig. 3.** Reconstructed phase noise from experimental PSD masks: (a) Power spectrum; (b) Oscillation phase in the time domain; (c) Instantaneous phase jitters extracted through a smoothing filter; (d) Probability density function.

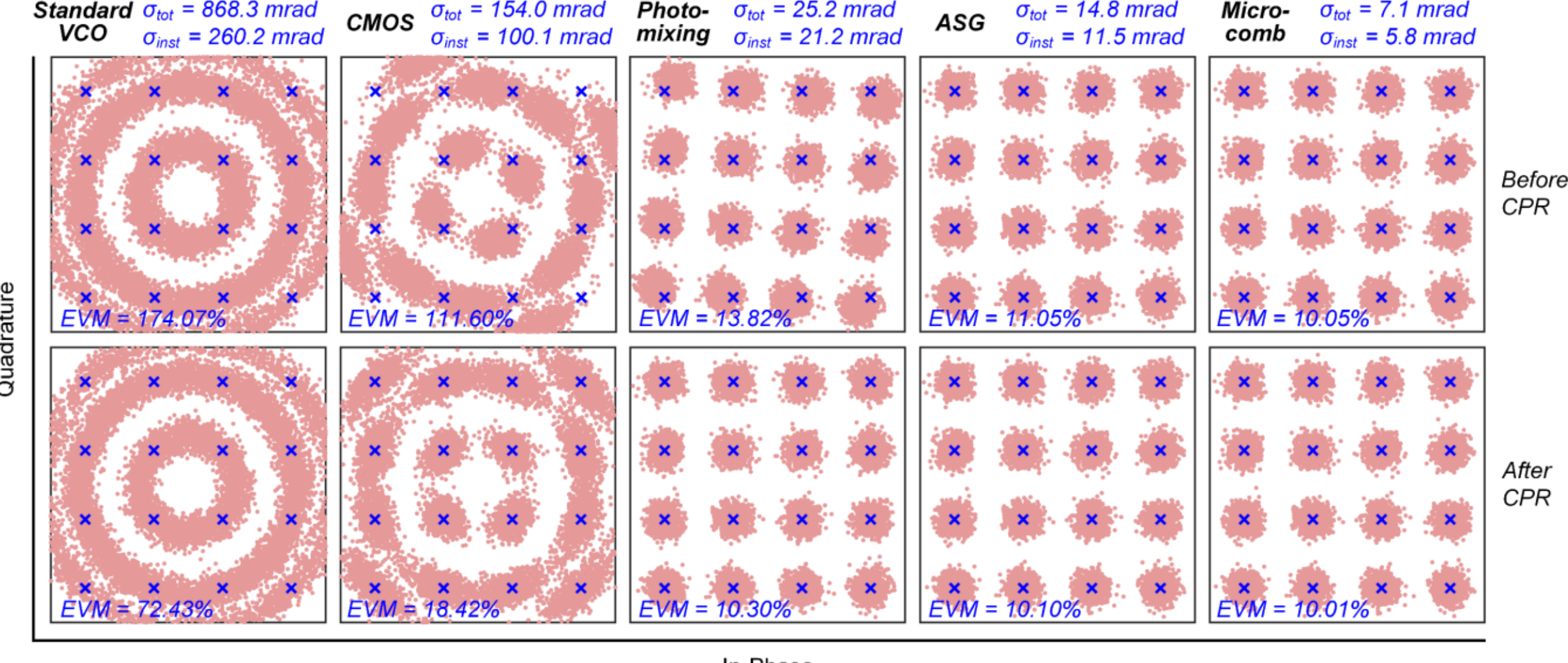

**Fig. 4.** Constellation diagrams of 16-QAM symbols driven by different THz sources before and after CPR under 20-dB SNR.

Therefore, the probability of a specific phase jitter is obtained by integrating the PDF over the corresponding interval.

$$P\left[\varphi_1 \le \varphi_{\text{inst}} \le \varphi_2\right] = \int_{\varphi_1}^{\varphi_2} f\left(\varphi\right) d\varphi \tag{13}$$

Figure 4 shows the constellations of received symbols driven by different THz sources before and after CPR under 20-dB SNR. As the phase noise strength decreases, the RMS EVM after CPR also declines from ~74% to 10%, as well as the dispersion of symbol mapping. For standard VCO and CMOS, excessive phase noise causes enormous rotation errors, even making circular constellation patterns. More importantly, two rotation patterns are induced by phase noise on the constellations: globally shared common phase error and locally dispersed instantaneous phase error. Processed with CPR, the common phase errors are eliminated, spinning the constellations back to their normal positions. Whereas instantaneous phase errors remain uncorrected, which decides the eventual transmission performance. This is consistent with the phase error mechanism. Low-offset phase noise leads to the inter-symbol rotation that all constellation coordinates share, while the high-offset component only introduces intra-symbol dispersion that surrounds their RMS center. It can be summarized that the low-offset phase noise is closely time-related and therefore could be suppressed to very low levels after PLL-assisted CPR. However, the high-offset component is greatly randomized, thus it is much challenging to correct. Consequently, the instantaneous uncertainty of the residential phase noise becomes an actual obstacle that constrains the error probability of high-order QAM traffic to a worse state, especially in a high-SNR scenario. This highlights the advantages of microcombs in high-throughput links.

Furthermore, Fig. 5 illustrates the CPR-processed constellations of 64-QAM and 256-QAM under 30-dB SNR driven by photonic-mixing ECL, ASG, and microcomb, respectively. From ECL to microcomb, along with the phase noise falling, the symbol rotation error is gradually eliminated, and the resulting EVM also declines from ~4% to 3.2%. The improvement in QAM ary does not cause additional growth of error magnitude, but the tolerance becomes more stringent due to the denser symbol mapping. More severely, high-ary QAM format indicates high symbol energy, which zooms in on the effect of phase noise, especially for edge mapping. Consequently, even a slight increase in phase error would seriously degrade the transmission performance for high-ary QAM mapping. This further enhances the benefits of adopting a high-$Q$ MRR based microcomb as the carrier source in THz wireless communications.

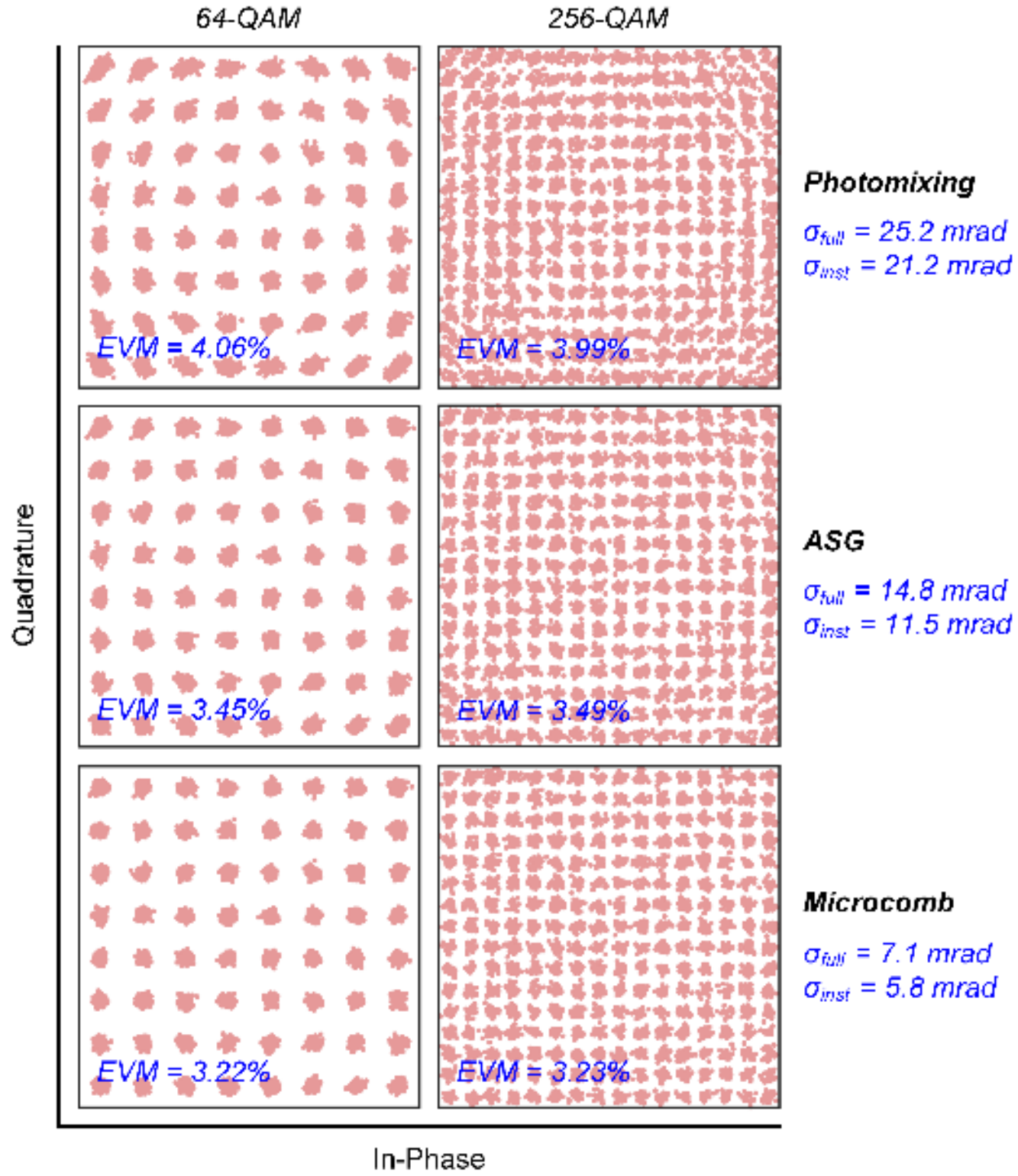

**Fig. 5.** CPR-processed constellations driven by different THz sources for 64-QAM and 256-QAM symbols under 30-dB SNR.

*B. BER Evaluation and Tolerance*

To quantitatively evaluate the impact of phase noise, BER is calculated first. Here, we assume a frequency-flat link without PM-AM conversion in devices and an AWGN channel. Given that the Monte Carlo method consumes massive computing resources but has moderate accuracy in the low BER region, the error probability model under a Gaussian channel is analyzed. Only the residual white phase noise needs to be addressed, because PLL-CPR eliminates most of the colored noise, which conforms to the normal distribution, like the intensity noise of the Gaussian channel. Therefore, the residual phase error per symbol has zero expected value and variance $\sigma_{\text{inst}}$, independent of the AWGN $n_0$.

$$y = r\cdot e^{-j\lambda} \approx u\cdot e^{j(\varphi-\mathbb{E}[\varphi])} + n_0 = u\cdot e^{j\varphi_{\text{inst}}} + n_0 \tag{14}$$

Making the small-angle linearization and keeping the first order as,

$$\text{for } \left|\varphi_{\text{inst}}\right| \ll 1 \text{ , } e^{j\varphi_{\text{inst}}} \approx 1 + j\varphi_{\text{inst}} \tag{15}$$

$$y(n) = u(n) + u(n)\cdot j\varphi_{\text{inst}}(n) + n_0 \tag{16}$$

Thus, the phase noise induced term is tangential to the constellation and orthogonal to the $n$-th QAM symbol $u(n)$. Given that the symbol sequence and phase jitters are independent. Its power $P_{\text{pn}}$ is,

$$\begin{aligned} P_{pn} &= \mathbb{E}\left[\left|u(n)\cdot j\varphi_{\text{inst}}(n)\right|^2\right] = \mathbb{E}\left[\left|u(n)\right|^2\right]\cdot\mathbb{E}\left[\left|\varphi_{\text{inst}}(n)\right|^2\right] \\ &= E_s\sigma_{\text{inst}}^2 \end{aligned} \tag{17}$$

Hence, phase noise does not degrade the original channel SNR directly—it adds an orthogonal error that reduces the Euclidean decision distance. Treating $ju(n)\varphi_{\text{inst}}(n)$ as an additive and independent noisy term, the total noise power during one symbol at the detector can be derived from,

$$N_{\text{tot}} = E_s\sigma_{\text{inst}}^2 + N_0 \tag{18}$$

Consequently, the effective SNR under impact of phase noise for error decision is calculated as follows,

$$SNR = \frac{E_s}{N_0} = \frac{\log_2 M \times E_b}{N_0} \tag{19}$$

$$SNR_{\text{eff}} = \frac{E_s}{N_{\text{tot}}} = \frac{E_s}{P_{\text{pn}} + N_0} = \frac{SNR}{1 + SNR\cdot\sigma_{\text{inst}}^2} \tag{20}$$

$$\begin{aligned} P_b &\approx \frac{4}{\log_2 M}\left(1 - \frac{1}{\sqrt{M}}\right) Q\left(\sqrt{\frac{3\log_2 M}{M-1} SNR_{\text{eff}}}\right) \\ &\text{where } Q(x) = \frac{1}{\sqrt{2\pi}}\int_x^{\infty} e^{-t^2/2}dt \end{aligned} \tag{21}$$

Where $M$ is the ary number, $E_b$, and $E_s$ are the average energy per bit and per symbol, respectively. This effective SNR reduces to the standard channel SNR when the impact of phase noise is eliminated as $\sigma_{\text{inst}} \rightarrow 0$. This mechanism yields a square $M$-QAM (Gray) equation under varying phase noise effects.

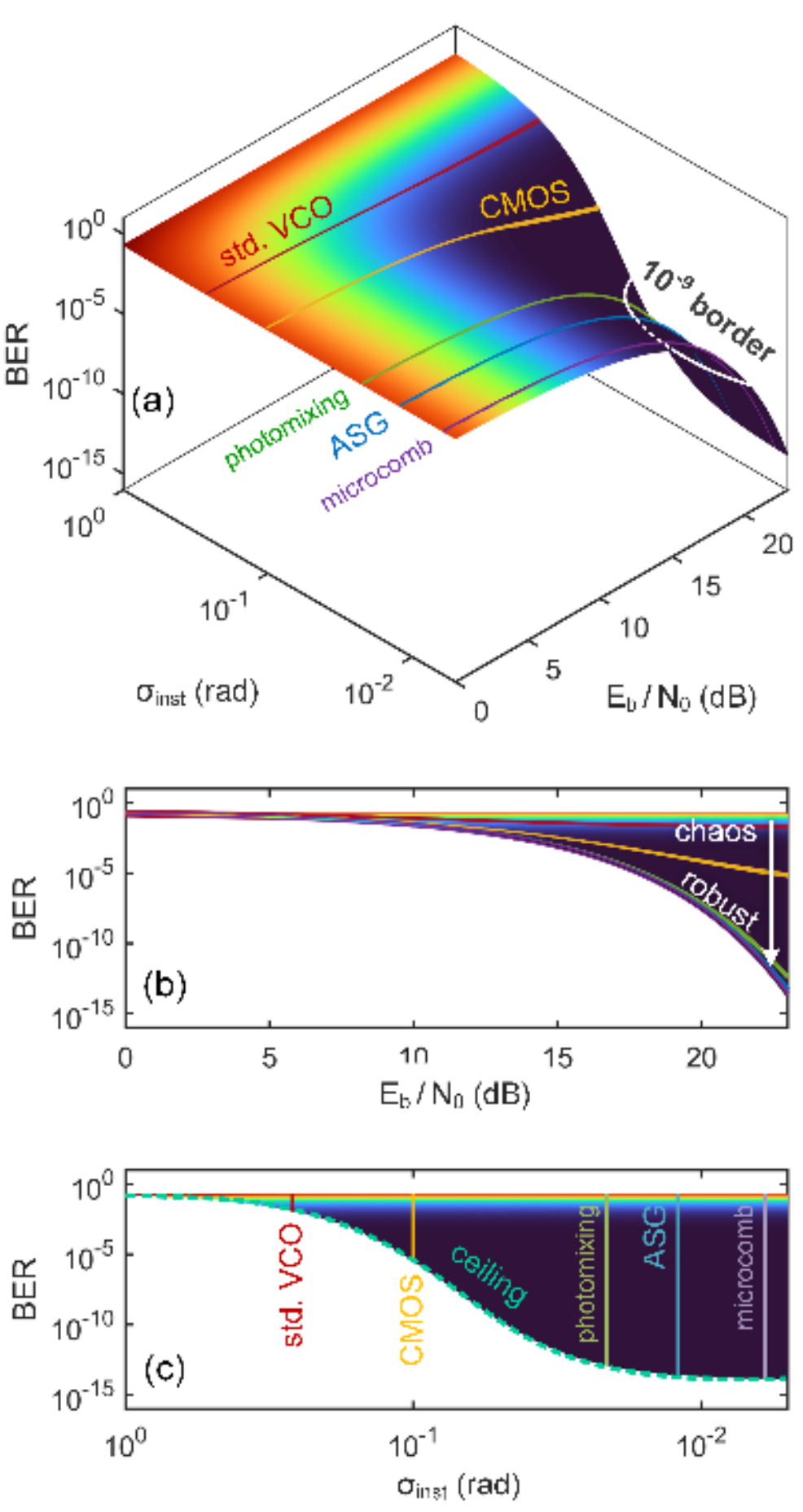

**Fig. 6.** BER evaluation for 64-QAM symbols: (a) Relationship between BER, ENR, and white phase noise; (b) Projection on the ENR-BER coordinate plane; (c) Projection on the $\sigma_{\text{inst}}$-BER coordinate plane.

Figure 6 predicted the 64-QAM BER surface versus phase noise magnitude and SNR per bit. As either the bit energy-to-noise ratio (ENR) $E_b/N_0$ increases or the residual phase noise amplitude $\sigma_{\text{inst}}$ decreases, the BER rapidly improves by several orders of magnitude. The descending trajectories corresponding to five THz oscillators are also marked in Fig. 6(a), highlighting the different error performance imposed by their phase noise levels. The projection in Fig. 6(b) maps the BER variation onto the ENR-BER coordinate plane. For an oscillator that suffers from strong phase noise, such as a standard VCO or a CMOS source, the corresponding BER curves lie in the upper chaotic region. Their slopes are shallow, and the sensitivity to ENR enhancement is significantly reduced. This implies a severe channel SNR penalty: at the same transmit power level, the final BER performance is much worse; conversely, achieving a target BER requires a substantially higher transmit power. In contrast, low-noise oscillators, such as a microcomb, approach the robust bound, where BER curves exhibit very steep slopes, indicating that even a small increase in transmit power could yield a pronounced BER improvement. Figure 6(c) provides a complementary projection in the $\sigma_{inst}$-BER coordinate plane, which reveals the intrinsic phase noise constraint under an ideal, infinitely high SNR condition. As $\sigma_{inst}$ decreases from high to low values, the achievable minimum BER transitions from a slowly varying plateau to a sharp decline and eventually saturates at the noise-free ceiling (~$1\times10^{-14}$). This indicates the existence of phase noise tolerance: once an oscillator's phase uncertainty is suppressed below this threshold, the related communication system can fully exploit the available SNR, reaching a state of energy-efficient transmission. Such noise tolerance is of critical importance for high-throughput wireless links and applications in AI networks and data center interconnections. The advantage of the microcomb is clearly reinforced by the lowest BER boundary.

Figure 7 generalizes the conclusions of Fig. 6 to different QAM formats. At 20-dB ENR, the correlation between BER and $\sigma_{inst}$ is summarized in Fig. 7(a). All the *M*-QAM exhibit a similar pattern of BER. It decreases slowly at high phase noise levels, then drops sharply as $\sigma_{inst}$ is reduced, and finally saturates at its ceiling value. This three-stage pattern reflects the unique dynamics of phase noise. At the high-noise stage, excessive phase uncertainty dominates the transmission errors even under a high SNR, thus preventing BER optimization. During the intermediate stage, error performance improves efficiently as phase noise is reduced, providing a high cost-performance margin for error optimization. In the quiet region below the phase noise tolerance, the lowest BER is no longer limited by phase noise itself, but instead constrained by the available SNR, or equivalently, transmit energy. The slope of the BER decrease also depends strongly on the QAM order. Higher-ary QAM formats (*e.g.* 128-QAM, 256-QAM) show shallower slopes and larger BER floors at low $\sigma_{inst}$, due to the narrower decision boundaries of dense constellation mapping. This agrees with the inherent sensitivity of high-order modulation to phase errors. The inset of Fig. 7(a) magnifies the low-noise region $\sigma_{inst} \in [0, 0.1]$ rad. It shows that microcomb, ASG, and ECL schemes support BER levels below $1\times10^{-9}$ for 64-QAM direct transmission and even enable 256-QAM links when protected by a 20% low-density parity-check (LDPC) code. For a stricter decision, BER<$1\times10^{-12}$, only the microcomb approach can secure reliable performance for 32-QAM transmission.

Furthermore, Fig. 7(b) summarizes the error-free borders of *M*-QAM transmission under combined intensity and phase noise conditions, assuming a BER threshold of $1\times10^{-9}$. In the large-noise region, the boundaries of different QAM orders diverge widely, resulting in steep terrain separation. This corresponds to a severe energy penalty—links driven by noisy oscillators require disproportionately high transmit power to achieve the same BER target. By contrast, when the phase noise magnitude is reduced into the tolerance region, higher-ary QAM formats become feasible with modest SNR budgets. This mechanism allows efficient utilization of transmit bit-energy, enabling a step-up in QAM order without prohibitive SNR costs.

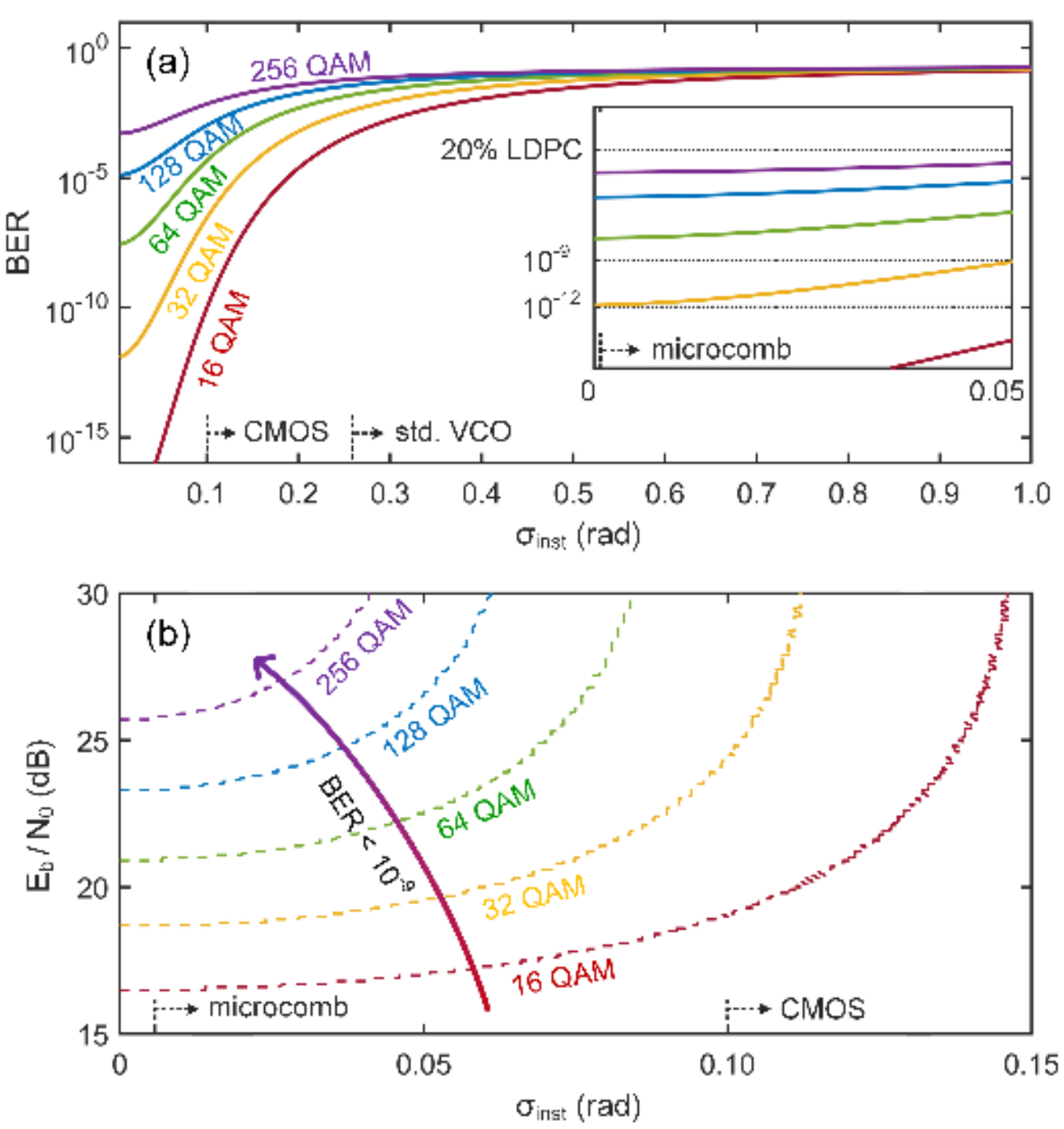

**Fig. 7.** (a) BER versus phase noise magnitude for *M*-QAM symbols under 20-dB ENR; (b) *M*-QAM boundary lines of the BER <$1\times10^{-9}$ decision region under varied noise.

## III. Analysis of Phase Noise tolerance and SNR Requirements

### *A. 3σ Error Criterion Based on the RMS-EVM*

To establish a simple tolerance rule, we adopt a $3\sigma$ error criterion based on the RMS-EVM. This criterion could directly connect the statistical distribution of phase noise to the constellation geometry from the Monte Carlo validator, providing a source-agnostic and modulation-order–specific tolerance threshold. Meanwhile, the $3\sigma$ rule is a conservative approach. By ensuring that nearly all symbol deviations remain within the decision region, the resulting BER lies far below typical pre-FEC limits. As illustrated in Fig. 8(a), residual instantaneous phase jitters $\varphi_{inst}$ cause azimuthal dispersion of QAM mapping. Accordingly, the constellation RMS-EVM depends on the magnitude of white phase noise $\sigma_{inst}$. Here, the concept of the PauTa criterion is borrowed to circle a high-probability region of hard-decision correctness under noise distortion. Thus, 99.7% of the samples lie within $\pm3\sigma$ coverage, which provides a conservative boundary for reliable symbol

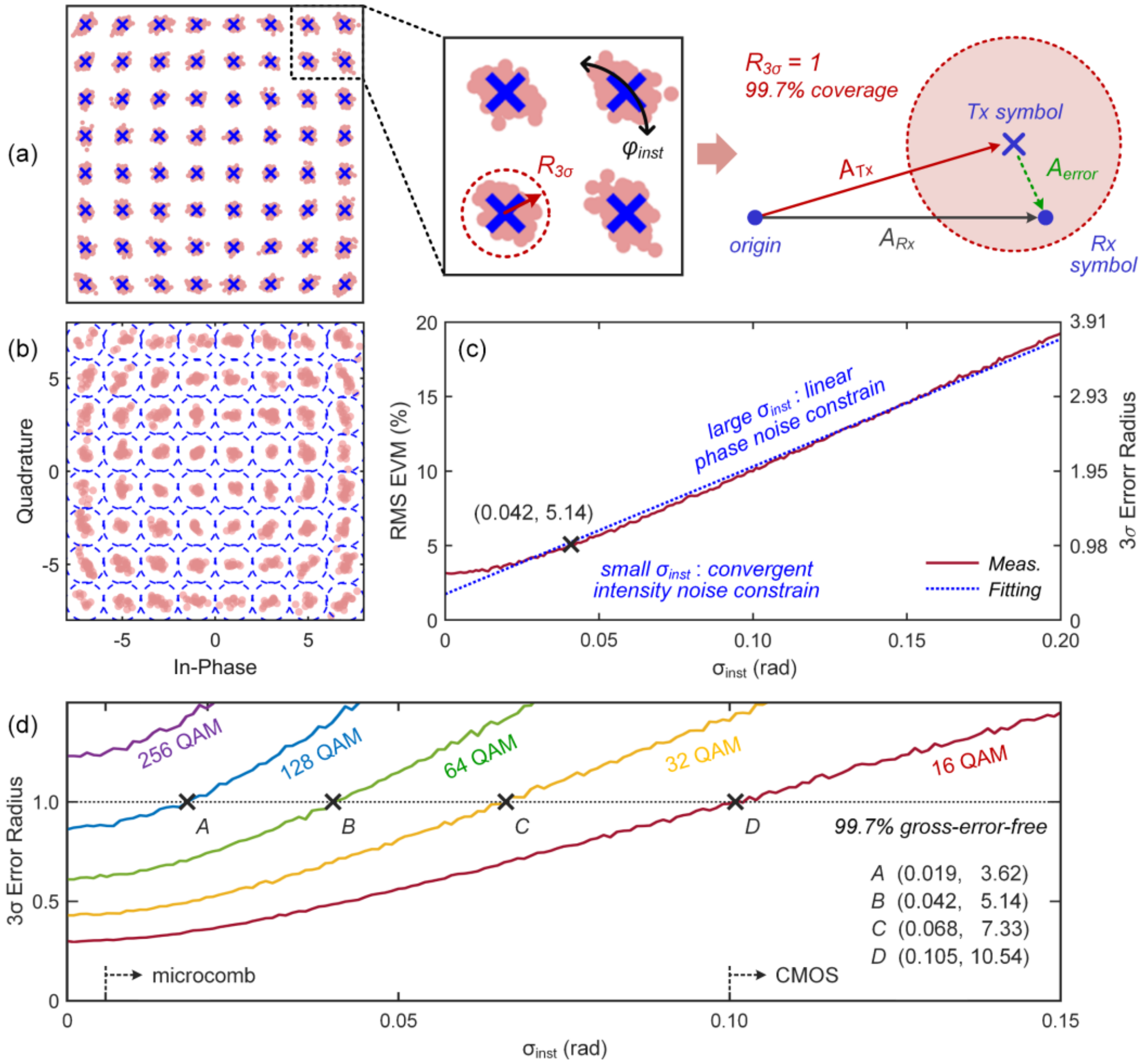

**Fig. 8.** (a) EVM estimation and its 3σ error coverage; (b) 64-QAM constellation when the 3$\sigma$ error radius $R_{3\sigma}$ = 1; (c) Linear correlation between RMS EVM (or $R_{3\sigma}$) and phase noise on the 64-QAM constellation; (d) 3$\sigma$ error radius versus phase noise magnitude for $M$-QAM.

detection.

For square $M$-QAM constellations, the minimum Euclidean distance between adjacent symbols is,

$$d_{\min} = \sqrt{\frac{6E_s}{(M-1)}} \tag{22}$$

The decision boundary is therefore located at half this distance. To satisfy the ±3$\sigma$ criterion for gross-error-free transmission, its radius $R_{3\sigma} \leq 1$ is referred to as the threshold.

$$R_{3\sigma} = 3 \times EVM_{\text{rms}} \cdot \sqrt{E_s} \leq \frac{1}{2} d_{\min} \tag{23}$$

Accordingly, the maximum tolerable RMS-EVM is,

$$EVM_{\text{rms}}(3\sigma) \leq \frac{1}{\sqrt{6(M-1)}} \tag{24}$$

which results in approximately 10.54% for 16-QAM, 5.14% for 64-QAM, and 2.56% for 256-QAM gross-error-free links. Figure 8(b) presents the 64-QAM constellation at a critical point with the 3$\sigma$-radius equal to one; that is, the RMS-EVM is 5.14% when $\sigma_{\text{inst}}$ is 42 mrad. Figure 8(c) shows the relationship between RMS-EVM and the magnitude of residual phase noise on a 64-QAM constellation diagram. Similarly, the RMS-EVM converges to the limit decided by SNR in the quiet region of low phase noise. As the phase noise magnitude increases, the RMS-EVM shows a linear positive correlation, which reflects that the phase jitter has become the main determinant of transmission error. For the 64-QAM format, the $\sigma_{\text{inst}}$ threshold for $R_{3\sigma}$ = 1 (*i.e.,* $EVM_{\text{rms}}$ = 5.14%) is 0.042 rad. This linear correlation is further generalized to $M$-QAM formats and is concluded in Fig. 8(d). The lower the white phase noise of a THz oscillator, the higher the QAM order it allows for direct transmission without gross errors. The tolerance of phase noise under a 3$\sigma$ error criterion, respectively, for 16-, 32-, 64-, and 128-QAM formats is 105 mrad, 68 mrad, 42 mrad, and 19 mrad (i.e., A, B, C, and D coordinates). As such, the 3$\sigma$ error criterion serves as a practical engineering guideline to assess whether a given THz source can support reliable transmission of a particular QAM format.

*B. Joint Analysis of Phase Noise Tolerance and SNR under 3σ Error Criterion*

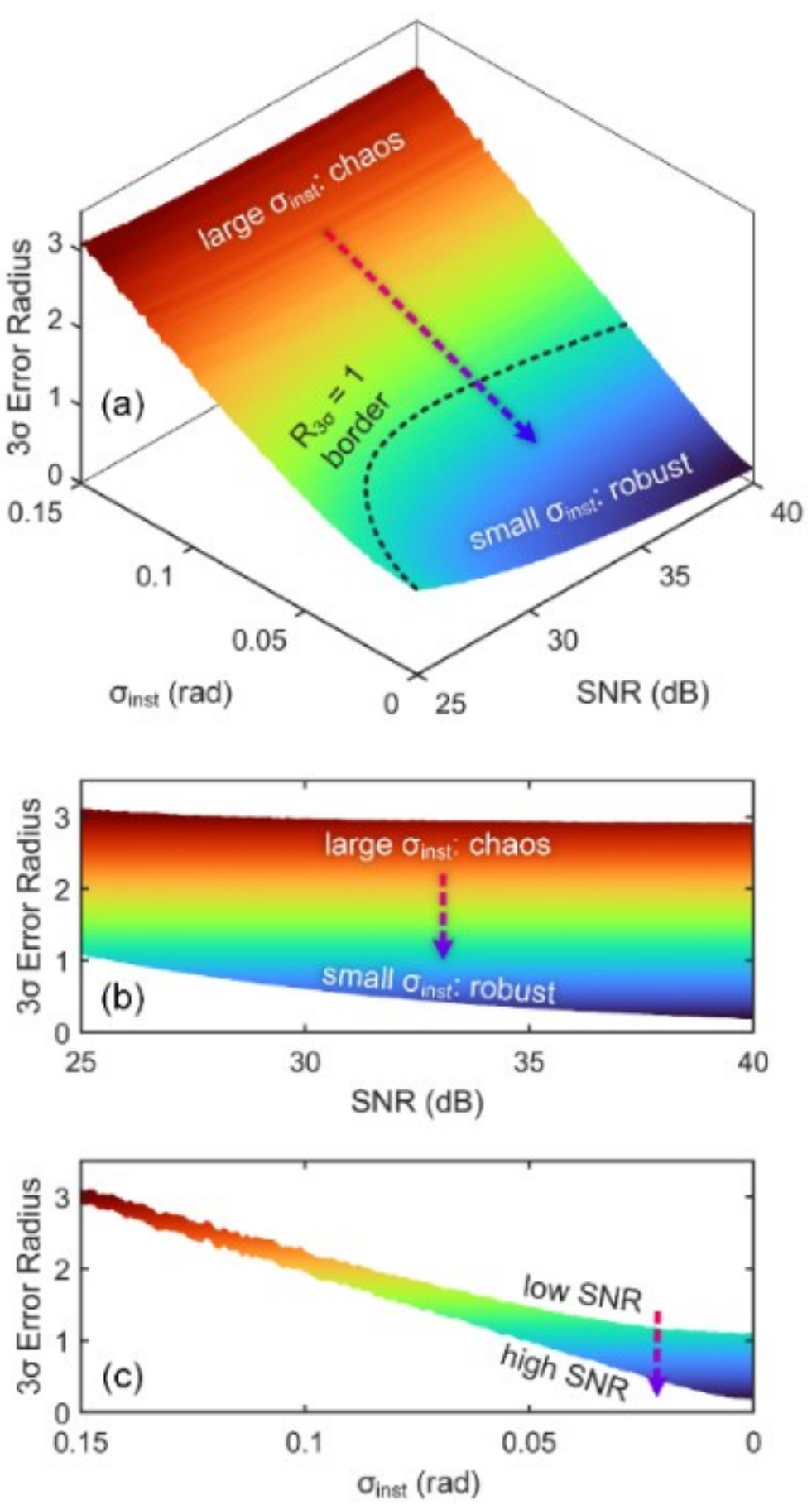

**Fig. 9.** $3\sigma$ error radius evaluation for 64-QAM symbols: (a) Relationship between radius, SNR, and white phase noise; (b) Projection on the radius-SNR coordinate plane; (c) Projection on the radius-$\sigma_{\text{inst}}$ coordinate plane.

Figure 9 presents the simulation results of the $3\sigma$ error radius under the combined impacts of the phase noise and the channel AWGN for 64-QAM. A smaller radius indicates tighter clustering of received symbols around their ideal positions, and hence a lower probability of detection errors. In Fig. 9(a), the most robust region is located at the far-right corner, corresponding to high SNR and small phase uncertainty, where the error radius is minimized. This agrees with the intuitive expectation that strong signal power and clean THz-oscillations jointly optimize wireless transmission fidelity. The projection in Fig. 9(b) shows the variation of $R_{3\sigma}$ with SNR for different levels of phase noise magnitude. The upper part of the map represents systems dominated by excessive phase noise, while the lower part corresponds to nearly noiseless oscillators. The sensitivity of $R_{3\sigma}$ to SNR differs markedly. When phase noise is negligible, the $3\sigma$ error radius decreases rapidly with increasing SNR, which is the desirable operating mode of a communication system. In contrast, under strong phase noise, $R_{3\sigma}$ becomes nearly insensitive to SNR, implying that additional transmit power does not translate into improved symbol reliability. Quantitatively, in the SNR range of 25~40 dB, the $3\sigma$ error radius under $\sigma_{\text{inst}} = 0.15$ rad can be 3 to 17 times larger than that in the ideal noiseless case. Figure 9(c) further highlights the dependence of $R_{3\sigma}$ on phase noise magnitude. Unlike the influence of SNR, reducing the white phase noise yields a much more pronounced improvement in error performance. This again emphasizes that oscillator quality in terms of low phase noise is the dominant factor in enabling robust high-order QAM THz transmission. Microcombs propellered by high-quality MRRs provide a decisive advantage by suppressing the noise floor well below the tolerance.

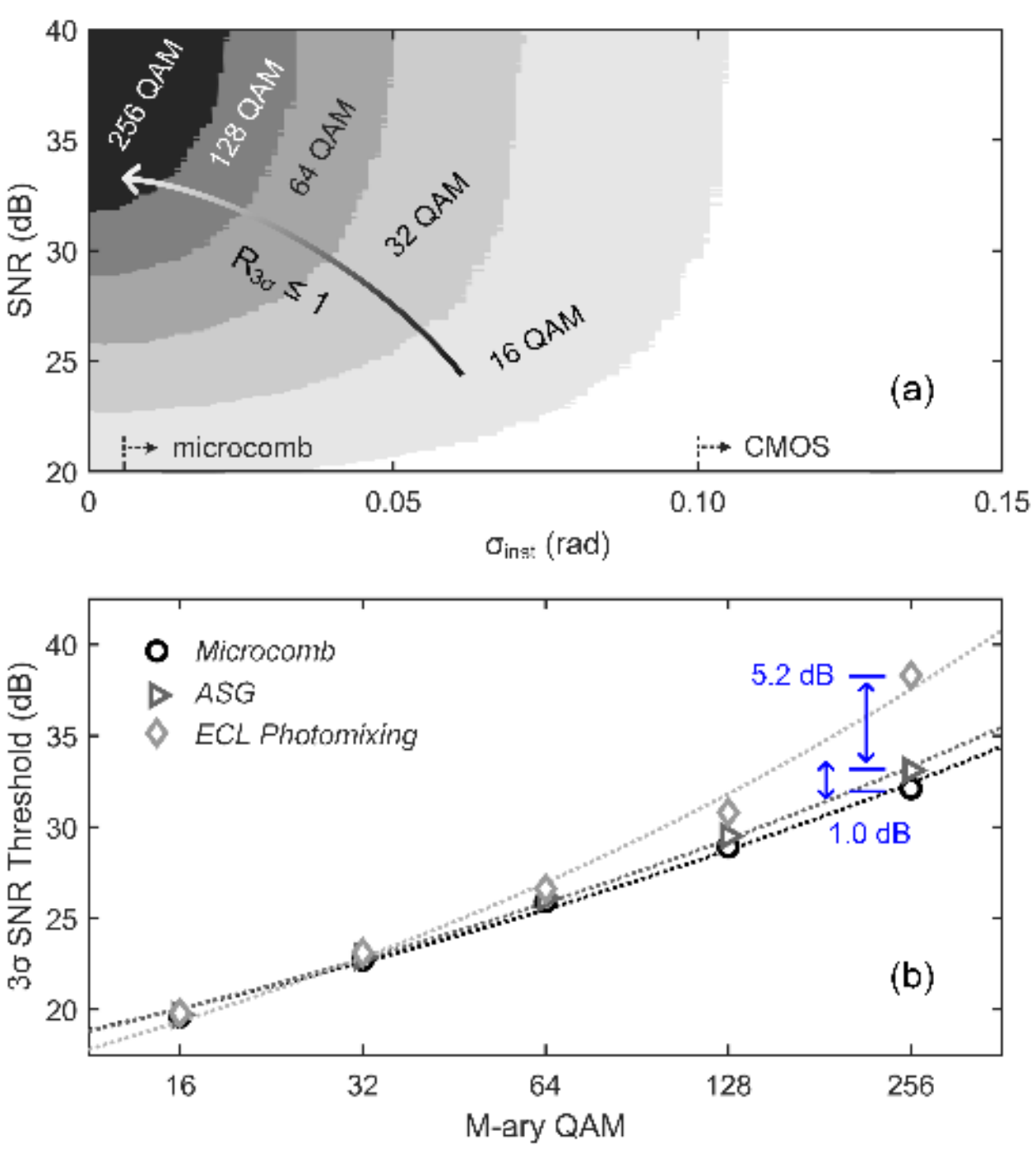

**Fig. 10.** (a) $3\sigma$ error decision regions for $M$-QAM symbols under varied noise; (b) SNR thresholds of $3\sigma$ error decision for $M$-QAM enabled by different THz-wave sources.

Moreover, Fig. 10 summarizes the tolerance borders extracted from Fig. 9(a) by taking the contour line of $R_{3\sigma} = 1$. The results for different QAM orders are compiled in Fig. 10(a). Along with the increasing ary number $M$, the feasible parameter space shrinks dramatically since high-order QAM requires both high SNR and low phase noise to maintain symbol reliability. This reveals the decisive role of oscillator phase noise. Suppressing the white phase-noise floor below the tolerance region is highly cost-effective, enabling no-FEC higher-order QAM transmission with minimal SNR overhead. By contrast, in the highly noisy region, the terrain of the link budget becomes prohibitively steep, representing a dead zone where stable transmission cannot be guaranteed even with infinite SNR. In extreme cases, even low-ary schemes such as 16-QAM become infeasible. These phase noise dynamics are further quantified in Fig. 10(b), where three THz sources with close phase noise performance—microcomb, ASG, and ECL schemes—are compared regarding the additional SNR overhead required to support different QAM orders. The results remarkably show that oscillators with higher residual phase noise suffer greater SNR penalties, and the penalty grows with the modulation order. For instance, in 256-QAM transmission, the SNR threshold difference between microcomb and ASG reaches ~1 dB, and that between microcomb and ECL reaches 6.2 dB, respectively. Such penalties on SNR are critical in practical deployment scenarios, where the link budget is already constrained by severe path loss and fluctuating channel environments. These results emphasize the importance of developing low-phase-noise THz oscillators by photonic approaches, as even moderate improvements yield substantial gains in spectral efficiency and transmission robustness.

## V. Conclusion

In summary, this paper focuses on the impact of oscillator phase noise on uncoded *M*-QAM SC-THz direct transmission. The symbol-distortion dynamics are revealed, and the tolerance of phase noise for different QAM orders is quantitatively explored. We reconstructed time-domain phase noise from measured spectra and embedded it in a simplified model of THz links. Distinguished rotation error patterns are identified, which are derived from slow CPE and instantaneous phase jitters, respectively. More importantly, CPE can be almost eliminated through the PLL-assisted CPR, while the instantaneous component cannot; therefore, the white phase uncertainty as a residual noise essentially dominates the system error performance. Furthermore, our simulation unveils a clear tolerance region regarding phase noise. Through suppressing the oscillator's noise floor into this region, modest increases in signal power could sharply push BER to new lows and enable high-order QAM. Beyond the threshold, however, reliable transmission cannot be guaranteed regardless of the available SNR. To intuitively quantify this effect, we proposed a $3\sigma$ error criterion, which shows a linear correlation between the EVM and the phase noise magnitude. The resulting phase-noise tolerances are approximately 105 mrad, 68 mrad, 42 mrad, and 19 mrad for 16-, 32-, 64-, and 128-QAM, respectively. Signal power penalty effect is also observed; for instance, the extracted tolerance of 256-QAM reflects an SNR penalty of ~6.2 dB between MRR-microcomb and ECL-photomixing schemes. These results reveal the dynamics of phase noise and define the robust operating region of *M*-QAM THz links, providing practical guidance for the design of the physical layer and protocols. The advantage of low-phase-noise photonic oscillators, such as microcomb, is significantly highlighted in enabling energy-efficient and spectrally efficient THz wireless communications.

## Acknowledgment

The authors would like to acknowledge (leave a blank) …

## References

[1] Cisco, "Cisco Annual Internet Report (2018–2023) White Paper," 2020, Accessed on: Sept. 12, 2021, [Online]. Available: https://www.cisco.com/c/en/us/solutions/collateral/executive-perspectives/annual-internet-report/white-paper-c11-741490.html

[2] K. Liu, Y. Feng, C. Han, B. Chang, Z. Chen, Z. Xu, L. Li, B. Zhang, Y. Wang, and Q. Xu, "High-speed 0.22 THz communication system with 84 Gbps for real-time uncompressed 8K video transmission of live events," Nat. Commun., vol. 15, no. 1, Sept. 2024, Art. no. 8037, doi: 10.1038/s41467-024-52370-x.

[3] Z. Chen, X. Ma, B. Zhang, Y. Zhang, Z. Niu, N. Kuang, W. Chen, L. Li, and S. Li, "A survey on terahertz communications," China Commun., vol. 16, no. 2, pp. 1–35, Feb. 2019.

[4] Y. Mochida, D. Shirai, and K. Takasugi, "Ultra-low-latency 8K-video-transmission system utilizing whitebox transponder with disaggregation configuration," IEICE Trans. Electron. or Trans. Electron., vol. E106-C, no. 6, pp. 321–330, Jun. 2023, doi: 10.1587/transele.2022LHP0003.

[5] H. Shams and A. Seeds, "Photonics, fiber and THz wireless communication," OPN, vol. 28, no. 3, pp.24–31, Mar. 2017, doi: 10.1364/OPN.28.3.000024.

[6] T. Nagatsuma, "Terahertz technologies: present and future," IEICE Electron. Express., vol. 8, no. 14, pp. 1127–1142, Jul. 2011, doi: 10.1587/elex.8.1127.

[7] M. Fujishima, "Overview of sub-terahertz communication and 300GHz CMOS transceivers," IEICE Electron. Express., vol. 18, no. 8, pp. 20212002, Apr. 2021, doi: 10.1587/elex.18.20212002.

[8] 3GPP TR 38.803, "Study on new radio access technology: radio frequency (RF) and co-existence aspects," ver. 14.4.0, Release 14, Accessed on: Jul. 10, 2024, [Online]. Available: https://www.3gpp.org/ftp/Specs/archive/38_series/38.803/38803-e40.zip.

[9] X. Lin, D. Yu, and H. Wiemann, "A primer on bandwidth parts in 5G new radio," in *5G and Beyond: Fundamentals and Standards*. Cham, Switzerland: Springer Nature Switzerland AG, 2021, pp. 357–370, doi: 10.1007/978-3-030-58197-8_12.

[10] S. Koenig, D. Lopez-Diaz, J. Antes, F. Boes, R. Henneberger, A. Leuther, A. Tessmann, R. Schmogrow, D. Hillerkuss, R. Palmer, T. Zwick, C. Koos, W. Freude, O. Ambacher, J. Leuthold, and I. Kallfass, "Wireless sub-THz communication system with high data rate," Nat. Photon., vol. 7, no. 12, pp. 977–981, Oct. 2013, doi: 10.1038/nphoton.2013.275.

[11] T. Ishibashi, Y. Muramoto, T. Yoshimatsu, and H. Ito, "Unitraveling-carrier photodiodes for terahertz applications," IEEE J. Sel. Top. Quantum Electron., vol. 20, no. 6, Jul. 2014, Art. no. 3804210, doi: 10.1109/JSTQE.2014.2336537.

[12] H. Ito, S. Kodama, Y. Muramoto, T. Furuta, T. Nagatsuma, and T. Ishibashi, "High-speed and high-output InP/InGaAs uni-travelingcarrier photodiodes," IEEE J. Sel. Topics Quantum Electron., vol. 10, no. 4, pp. 709–727, Aug. 2004, doi: 10.1109/JSTQE.2004.833883.

[13] J. Ma, R. Shrestha, L. Moeller, and D. M. Mittleman, "Channel performance for indoor and outdoor terahertz wireless links," APL Photonics, vol. 3, no. 5, May 2018, Art. no. 051601, doi: 10.1063/1.5014037.

[14] T. Nagatsuma, K. Oogimoto, Y. Yasuda, Y. Fujita, Y. Inubushi, and S. Hisatake, "300-GHz-band wireless transmission at 50 Gbit/s over 100 meters," in *IRMMW-THz*, Copenhagen, Denmark, 2016, pp. 1–2, doi: 10.1109/IRMMW-THz.2016.7758356.

[15] X. Pang, A. Caballero, A. Dogadaev, V. Arlunno, R. Borkowski, J. S. Pedersen, L. Deng, F. Karinou, F. Roubeau, D. Zibar, X. Yu, and I. T. Monroy, "100 Gbit/s hybrid optical fiber-wireless link in the W-band (75-110 GHz)," Opt. Express, vol. 19, no. 25, pp. 24944–24949, Nov. 2011, doi: 10.1364/OE.19.024944.

[16] X. Yu, S. Jia, H. Hu, M. Galili, T. Morioka, P. U. Jepsen, and L. K. Oxenløwe, "160 Gbit/s photonics wireless transmission in the 300-500 GHz band," APL Photonics, vol. 1, no. 8, Nov. 2016, Art. no. 081301, doi: 10.1063/1.4960136.

[17] Y. Chen, H. Peng, D. Fang, J. Dittmer, G. Lihachev, A. Voloshin, S. T. Skacel, M. Lauermann, A. Tessmann, S. Wagner, S. Bhave, I. Kallfass, T. Zwick, W. Freude, S. Randel, T. J. Kippenberg, and C. Koos, "Self-injection-locked Kerr soliton microcombs with photonic wire bonds for use in terahertz communications," in *CLEO*, San Jose, CA, United States, 2023, pp. STh3J.1, doi: 10.1364/CLEO_SI.2023.STh3J.1.

[18] T. Harter, C. Füllner, J. N. Kemal, S. Ummethala, J. L. Steinmann, M. Brosi, J. L. Hesler, E. Bründermann, A.-S. Müller, W. Freude, S. Randel, and C. Koos, "Generalized Kramers–Kronig receiver for coherent terahertz communications," Nat. Photon., vol. 14, no. 10, pp. 601–606, Sept. 2020, doi: 10.1038/s41566-020-0675-0.

[19] W. Li, J. Yu, B. Zhu, F. Wang, J. Ding, J. Zhang, M. Zhu, F. Zhao, T. Xie, K. Wang, Y. Wei, X. Yang, B. Hua, M. Lei, Y. Cai, L. Zhao, W. Zhou, and J. Yu, "Photonic terahertz wireless communication: towards the goal

of high-speed kilometer-level transmission," J. Lightwave Technol., vol. 42, no. 3, pp. 1159–1172, Nov. 2023, doi: 10.1109/JLT.2023.3329351.

[20] K. Maekawa, T. Yoshioka, T. Nakashita, T. Ohara, and T. Nagatsuma, "Single-carrier 220-Gbit/s sub-THz wireless transmission over 214 m using a photonics-based system" Opt. Lett., vol. 49, no. 16, pp. 4666–4668, Aug. 2024, doi: 10.1364/OL.527593.

[21] D. Fang, H. Peng, Y. Chen, J. Dittmer, A. Tessmann, S. Wagner, P. Matalla, D. Drayss, G. Lihachev, A. Voloshin, S. T. Skacel, M. Lauermann, I. Kallfass, T. Zwick, W. Freude, T. J. Kippenberg, S. Randel, and C. Koos, "Wireless THz communications at 250 Gbit/s using self-injection-locked Kerr soliton microcombs as photonic-electronic oscillators at the transmitter and receiver," in *ECOC*, Frankfurt, Germany, 2024, pp. 558–561.

[22] T. Nagatsuma, W. Gao, Y. Kawamoto, T. Ohara, H. Ito, and T. Ishibashi, "InP-based integrated circuits on SiC/Si substrates for terahertz communications," J. Lightwave Technol., early access, doi: 10.1109/JLT.2025.3561665.

[23] K. Maekawa, T. Nakashita, T. Yoshioka, T. Hori, A. Rolland, and T. Nagatsuma, "Single-channel 240-Gbit/s sub-THz wireless communications using ultra-low phase noise receiver," IEICE Electron. Express, vol. 21, no. 3, pp. 20230584, Feb. 2024, doi: 10.1587/elex.20.20230584.

[24] S. Jia, L. Zhang, S. Wang, W. Li, M. Qiao, Z. Lu, N. M. Idrees, X. Pang, H. Hu, X. Zhang, L. K. Oxenløwe, and X. Yu "2 × 300 Gbit/s line rate PS-64QAM-OFDM THz photonic-wireless transmission," J. Lightwave Technol., vol. 38, no. 17, pp. 4715–4721, Sept. 2020, doi: 10.1109/JLT.2020.2995702.

[25] H. Zhang, Z. Yang, Z. Lyu, H. Yang, L. Zhang, O. Ozolins, X. Pang, X. Zhang, and X. Yu, "300 GHz photonic-wireless transmission with aggregated 1.034 Tbit/s data rate over 100 m wireless distance," in *OFC*, San Diego, CA, United States, 2024, pp. M2F.1, doi: 10.1364/OFC.2024.M2F.1.

[26] T Tetsumoto and A Rolland, "300 GHz wireless link based on whole comb modulation of integrated Kerr soliton combs," IEEE Photonics J., vol. 15, no. 6, Dec. 2023, Art. no. 7304809, doi: 10.1109/JPHOT.2023.3325088.

[27] Y. Geng, X. Han, X. Zhang, Y. Xiao, S. Qian, Q. Bai, Y. Fan, G. Deng, Q. Zhou, K. Qiu, J. Xu, and H. Zhou, "Phase noise of Kerr soliton dual microcombs," Optica, vol. 47, no. 18, pp. 4838–4841, Sept. 2022, doi: 10.1364/OL.469950.

[28] E. Lucas, P. Brochard, R. Bouchand, S. Schilt, T. Südmeyer, and T. J. Kippenberg, "Ultralow-noise photonic microwave synthesis using a soliton microcomb-based transfer oscillator," Nat. Commun., vol. 11, no. 1, Jan. 2020, Art. no. 374, doi: 10.1038/s41467-019-14059-4.

[29] W. Jin, Q. Yang, L. Chang, B. Shen, H. Wang, M. A. Leal, L. Wu, M. Gao, A. Feshali, M. Paniccia, K. J. Vahala, and J. E. Bowers, "Hertz-linewidth semiconductor lasers using CMOS-ready ultra-high-Q microresonators," Nat. Photon., vol. 15, no. 5, pp. 346–353, Feb. 2021, doi: 10.1038/s41566-021-00761-7.

[30] T. Tetsumoto, T. Nagatsuma, M. E. Fermann, G. Navickaite, M. Geiselmann, and A. Rolland, "Optically referenced 300 GHz millimetre-wave oscillator," Nat. Photon., vol. 15, no. 7, pp. 516–522, Apr. 2021, doi: 10.1038/s41566-021-00790-2.

[31] I. Kudelin, W. Groman, Q. Ji, J. Guo, M. L. Kelleher, D. Lee, T. Nakamura, C. A. McLemore, P. Shirmohammadi, S. Hanifi, H. Cheng, N. Jin, L. Wu, S. Halladay, Y. Luo, Z. Dai, W. Jin, J. Bai, Y. Liu, W. Zhang, C. Xiang, L. Chang, V. Iltchenko, O. Miller, A. Matsko, S. M. Bowers, P. T. Rakich, J. C. Campbell, J. E. Bowers, K. J. Vahala, F. Quinlan, and S. A. Diddams, "Photonic chip-based low-noise microwave oscillator," Nature, vol. 627, no. 8004, pp. 534–539, Mar. 2024, doi: 10.1038/s41586-024-07058-z.

[32] T. Wildi, A. E. Ulanov, T. Voumard, B. Ruhnke, and T. Herr, "Phase-stabilised self-injection-locked microcomb," Nat. Commun., vol. 15, no. 1, Aug. 2024, Art. no. 7030, doi: 10.1038/s41467-024-50842-8.

[33] A. E. Ulanov, T. Wildi, N. G. Pavlov, J. D. Jost, Ma. Karpov, and T. Herr, "Synthetic reflection self-injection-locked microcombs," Nat. Photon., vol. 18, no. 3, pp. 294–299, Jan. 2024, doi: 10.1038/s41566-023-01367-x.

[34] F. Lei, Z. Ye, Ó. B. Helgason, A. Fülöp, M. Girardi, and V. Torres-Company, "Optical linewidth of soliton microcombs," Nat. Commun., vol. 13, no. 1, Jun. 2022, Art. no. 3161, doi: 10.1038/s41467-022-30726-5.

[35] A. C. Triscari, A. Tusnin, A. Tikan, and T. J. Kippenberg, "Quiet point engineering for low-noise microwave generation with soliton microcombs," Commun. Phys., vol. 6, no. 1, Nov. 2023, Art. no. 318, doi: 10.1038/s42005-023-01437-0.

[36] D. B. Leeson, "A simple model of feedback oscillator noise spectrum," Proc. IEEE, vol. 54, no. 2, pp. 329–330, Feb. 1966, doi: 10.1109/PROC.1966.4682.

[37] P. Andreani and A. Bevilacqua, "Harmonic oscillators in CMOS—a tutorial overview," IEEE Open J. Solid-State Circuits Soc., vol. 1, pp. 2–17, Sept. 2021, doi: 10.1109/OJSSCS.2021.3109854.

[38] B. Liu and T. Tanabe, "Phase noise tolerance for low-pilot-overhead OFDM terahertz links beyond 64-QAM," in *ACP*, Suzhou, China, 2025, pre-print, doi: 10.48550/arXiv.2508.05026.

[39] B. Liu and T. Tanabe, "Impacts of phase noise on M-ary QAM THz wireless communications," in *IEEE OGC*, Shenzhen, China, 2025, pre-print, doi: 10.48550/arXiv.2508.01291.

[40] O. Memioglu, Y. Zhao, and B. Razavi, "A 300-GHz 52-mW CMOS receiver with on-chip LO generation," IEEE J. Solid-State Circuits., vol. 58, no. 8, pp. 2141–2156, Aug. 2023, doi: 10.1109/JSSC.2023.3257820.

[41] A. Franceschin, D. Riccardi, and A. Mazzanti, "Ultra-low phase noise X-band BiCMOS VCOs leveraging the series resonance," IEEE J. Solid-State Circuits., vol. 57, no. 12, pp. 3514–3526, Dec. 2022, doi: 10.1109/JSSC.2022.3202405.

[42] W. Wu, C. Yao, C. Guo, P. Chiang, L. Chen, P. Lau, Z. Bai, S. W. Son, and T. B. Cho, "A 14-nm ultra-low jitter fractional-n PLL using a DTC range reduction technique and a reconfigurable dual-core VCO," IEEE J. Solid-State Circuits., vol. 56, no. 12, pp. 3756–3767, Dec. 2021, doi: 10.1109/JSSC.2021.3111134.